\documentstyle[prd,aps,preprint,eqsecnum,psfig,floats]{revtex}

\def\bege{\begin{equation}}
\def\ende{\end{equation}}

\def\a{\alpha}
\def\b{\beta}

\def\L{\Lambda}

\def\del{\partial}

\def\beq{\begin{equation}}
\def\eeq{\end{equation}}
\def\bea{\begin{eqnarray}}
\def\eea{\end{eqnarray}}

\begin{document}
\tighten

\title{Renormalization-Scale-Invariant PQCD Predictions for
$R_{e^+e^-}$ and the Bjorken Sum Rule at Next-to-Leading Order}

\author{Michael Binger and Chueng-Ryong Ji}

\address{Department of Physics, North Carolina State University,
Raleigh, NC 27695-8202}

\author{David G. Robertson}

\address{Ohio Supercomputer Center, 1224 Kinnear Road, Columbus, OH
43212-1163}


\maketitle

\begin{abstract}
We discuss application of the physical QCD effective charge $\a_V$,
defined via the heavy-quark potential, in perturbative calculations at
next-to-leading order.  When coupled with the Brodsky-Lepage-Mackenzie
prescription for fixing the renormalization scales, the resulting
series are automatically and naturally scale and scheme independent,
and represent unambiguous predictions of perturbative QCD.  We
consider in detail such commensurate scale relations for the $e^+e^-$
annihilation ratio $R_{e^+e^-}$ and the Bjorken sum rule.  In both
cases the improved predictions are in excellent agreement with
experiment.

\end{abstract}

\section{Introduction}

One of the most important problems in making reliable predictions in
perturbative QCD is dealing with the dependence of the truncated
perturbative series on the choice of renormalization scale $\mu$ and
scheme $s$ for the QCD coupling $\alpha_s(\mu)$.  Consider a physical
quantity ${\cal O}$, computed in perturbation theory and truncated at
next-to-leading order (NLO) in $\alpha_s$:
\begin{equation}
{\cal O} = \a_s(\mu)\left[1+\left(A_1(\mu)+B_1(\mu) n_f\right)
{\a_s(\mu)\over\pi} + \cdots\right]\; ,
\label{1}
\end{equation}
where $n_f$ is the effective number of quark flavors.  The
finite-order expression depends on both $\mu$ and the choice of scheme
used to define the coupling.  In fact, Eqn. (\ref{1}) can be made to
take on essentially any value by varying $\mu$ and the renormalization
scheme, which are {\em a priori} completely arbitrary.  The
scale/scheme problem is that of choosing $\mu$ and the scheme $s$ in
an ``optimal'' way, so that an unambiguous theoretical prediction,
ideally including some plausible estimate of theoretical
uncertainties, can be made.\footnote{The precise meaning of
``optimal'' in this context is connected to the minimization of
remainders for the truncated series.  As is well known, perturbation
series in QCD are asymptotic, and thus there is an optimum number of
terms that should be computed for a given observable.  In general,
very little is known about the remainders in pQCD; however, if we
assume that pQCD series are sign-alternating, then the remainder can
be estimated by the first neglected (or last included) term.  This
term can take on essentially any value, however, by simply varying the
scale and scheme, and thus its minimization is meaningless without
invoking additional criteria.}

For any given observable there is no rigorously correct way to make
this choice in general.  However, a particular prescription may be
supported to a greater or lesser degree by general theoretical
arguments and, {\em a posteriori}, by its success in practical
applications.  From these perspectives, a particularly successful
method for choosing the renormalization scale is that proposed by
Brodsky, Lepage and MacKenzie \cite{BLM}.  In the BLM procedure, the
renormalization scales are chosen such that all vacuum polarization
effects from fermion loops are absorbed into the running couplings.  A
principal motivation for this choice is that it reduces to the correct
prescription in the case of Abelian gauge theory.  Furthermore, the
BLM scales are physical in the sense that they typically reflect the
mean virtuality of the gluon propagators.  Another important advantage
of the method is that it ``pre-sums" the large and strongly divergent
terms in the pQCD series which grow as $n!  (\alpha_s \beta_0 )^n$,
i.e., the infrared renormalons associated with coupling constant
renormalization.

Dependence on the renormalization scheme can be avoided by considering
relations between physical observables only.  By the general
principles of renormalization theory, such a relation must be
independent of any theoretical conventions, in particular the choice
of scheme in the definition of $\alpha_s$.  A relation between
physical quantities in which the BLM method has been used to fix the
renormalization scales is known as a ``commensurate scale relation''
(CSR) \cite{CSR}.  An important example is the generalized Crewther
relation \cite{CSR,bkgl}, in which the radiative corrections to the
Bjorken sum rule for deep inelastic lepton-proton scattering at a
given momentum transfer $Q$ are predicted from measurements of the
$e^+ e^-$ annihilation cross section at a commensurate energy scale
$\sqrt s \propto Q$.

A useful tool in these analyses is the concept of an ``effective
charge.''  Any perturbatively calculable observable can be used to
define an effective charge by incorporating the entire radiative
correction into its definition.  Since such a charge is itself a
physical observable, perturbation theory in terms of it, with the BLM
prescription setting the scales, is automatically renormalization
scale- and scheme-independent.

In this paper we shall use the heavy quark potential to define an
effective QCD coupling $\a_V$, and construct scale-commensurate
expansions of various other QCD observables in terms of it.  A recent
calculation of the heavy quark potential at NNLO \cite{peter} allows
the relevant BLM scales to be determined through NLO.  The resulting
relations can be tested directly for agreement with available data,
and in addition may be used to study various phenomenological forms
for the heavy quark potential at moderate to low $Q^2$.

We begin by outlining the BLM approach and the idea of commensurate
scale relations.  We also introduce physical effective charges
asociated with the heavy quark potential, the $e^+e^-$ annihilation
cross section and the Bjorken sum rule.  In section III we then
construct the NLO scale-commensurate expansions of these observables
in terms of $\a_V$, and compare the results to the available data
using a simple parameterization for $\a_V$ which is fit to a lattice
calculation.  In general the agreement is excellent.  In section IV we
present some discussion of the results and our conclusions.

\section{BLM Scale Fixing}

At lowest order the BLM approach is straightforward to motivate.  The
term involving $n_f$ in Eqn. (\ref{1}) arises solely from quark loops
in vacuum polarization diagrams.  In QED these are the only
contributions responsible for the running of the coupling, and thus it
is natural to absorb them into the definition of the coupling.  The
BLM procedure is the analog of this approach in QCD.  Specifically, we
rewrite Eqn. (\ref{1}) in the form
\begin{equation}
{\cal O} = \a_s(\mu)\left[1 -
\left({3\b_0 B_1(\mu)\over2}\right) {\a_s(\mu)\over\pi}
+\cdots\right]
\left[1 + 
\left(A_1(\mu)+ {33B_1(\mu)\over2}\right){\a_s(\mu)\over\pi}
+\cdots\right]\; ,
\end{equation}
correct to order $\a_s^2$, where $\b_0 = 11-{2n_f/3}$ is the
lowest-order QCD beta function.  The first term in square brackets can
then be absorbed by a redefinition of the renormalization scale in the
leading-order coupling, using
\begin{equation}
\a_s(\mu^*) = \a_s(\mu)\left[1-{\b_0\a_s(\mu)\over2\pi}
\ln\left(\mu^*/\mu\right)+ \cdots\right]\; .
\end{equation}
That is, the BLM procedure consists of defining the prediction for
${\cal O}$ at this order to be
\begin{equation}
{\cal O} = \a_s(\mu^*)\left[1 + \left(A_1(\mu)+ {33B_1(\mu)\over2} \right)
{\a_s(\mu^*)\over\pi}  + \cdots\right]\; ,
\end{equation}
where
\begin{equation}
\mu^* \equiv \mu e^{3B_1(\mu)}\; .
\end{equation}
Note that knowledge of the NLO term in the expansion is necessary to
fix the scale at LO.  Thus the scale occurring in the highest term in
the expansion will in general be unknown.  A natural prescription is
to set this scale to be the same as that in the next-to-highest-order
term.

A very important feature of this prescription is that $\mu^*$ is
actually independent of $\mu$.
(This follows from considering the $\mu$ dependence of $B_1(\mu)$.
For a detailed discussion of this point, see Ref. \cite{BLM}.)
Thus pQCD predictions using the BLM procedure are unambiguous.

The same basic idea can be extended to higher orders, by
systematically shifting $n_f$ dependence into the renormalization
scales order by order.  Full details of this procedure may be found in
Refs. \cite{bl95,luthesis}.  The result is that a general expansion
\begin{equation}
{\a_s(\mu)\over\pi} 
+ \left(A_1+B_1 n_f\right) \left({\a_s(\mu)\over\pi}\right)^2
+ \left(A_2+B_2 n_f + C_2 n_f^2\right) 
\left({\a_s(\mu)\over\pi}\right)^3 + \cdots
\end{equation}
is replaced by a series of the form
\begin{equation}
{\a_s(\mu^*)\over\pi} 
+ \widetilde{A}_1\left({\a_s(\mu^{**})\over\pi}\right)^2
+ \widetilde{A}_2\left({\a_s(\mu^{***})\over\pi}\right)^3
+ \cdots\; .
\end{equation}
In general a different scale appears at each order in perturbation
theory, and the BLM scales themselves are power series in the coupling
$\a_s$.  In addition, the coefficients $\widetilde{A}_n$ are
independent of $n_f$ (by construction), and so the form of the
expansion is unchanged as momenta vary across quark mass thresholds.
All effects due to quark loops in vacuum polarization diagrams are
automatically incorporated into the effective couplings.

As discussed above, one motivation for this prescription is that it
reduces to the correct result in the case of QED.  In addition, when
combined with the idea of commensurate scale relations, the BLM method
can be shown to be consistent with the generalized renormalization
group invariance of St\"uckelberg and Peterman \cite{sp53}, in which
one considers ``flow equations'' both in $\mu$ and in the parameters
that define the scheme \cite{bl95}.  This is not necessarily true of
other methods for determining the scales.

A very natural way of implementing the CSR idea is to introduce a
physical effective charge, defined via some convenient observable, for
use as an expansion parameter.  An expansion of a physical quantity in
terms of such a charge is a relation between observables and therefore
must be independent of theoretical conventions, such as the
renormalization scheme, to any fixed order of perturbation theory.  A
particularly useful scheme is furnished by the heavy quark potential
$V(Q^2)$, which can be identified as the two-particle-irreducible
amplitude for the scattering of an infinitely heavy quark and
antiquark at momentum transfer $t = -Q^2$.  The relation
\begin{equation}
V(Q^2) = - {4 \pi C_F \alpha_V(Q)\over Q^2}\; ,
\end{equation}
with $C_F=(N_c^2-1)/2 N_c=4/3$, then defines the effective charge
$\alpha_V(Q)$.  This coupling provides a physically-based alternative
to the usual ${\overline {MS}}$ scheme.  The other physical charges we
shall consider here are $\a_R$, defined via the total $e^+e^-\to X$
cross section:
\begin{equation}
R(s) \equiv 3\sum_q e_q^2 \left(1 + {\alpha_R(\sqrt s)\over
\pi}\right)\; ,
\label{alphardef}
\end{equation}
and $\a_{g_1}$, defined by the radiative correction to the Bjorken Sum
Rule:
\begin{equation}
\int_0^1 dx\left[g_1^{ep}(x,Q^2) - g_1^{en}(x,Q^2)\right]\equiv
{1\over6}\left| {g_A\over g_V}\right|
\left[1-{\a_{g_1}(Q)\over\pi}\right]\; .
\end{equation}
The perturbative expansions for these quantities through NNLO may be
found in Refs. \cite{rrefs} and \cite{bjrefs,aptbsr}, respectively.

Such physical couplings are of course renormalization-group-invariant,
i.e., $\mu\del \a_V/\del\mu = 0$.  However, the dependence of
$\a_V(Q)$ on $Q$ is controlled by an equation which is formally
identical to the usual RG equation.  Since $\a_V$ is dimensionless we
must have
\begin{equation}
\a_V = \a_V\left({Q\over\mu}, \a_s(\mu)\right)\; .
\end{equation}
Then $\mu\del\a_V/\del\mu = 0$ implies
\begin{equation}
Q{\del\over\del Q}\a_V(Q) = \b_s(\a_s){\del\a_V\over\del\a_s}
\equiv \b_V(\a_V)\; ,	
\end{equation}
where
\begin{equation}
\b_s = \mu{\del\over\del\mu}\a_s(\mu)\; .
\end{equation}
This is formally a change of scheme, so that the first two
coefficients $\beta_V^{(0)} = 11 - 2 n_f/3$ and $\beta_V^{(1)} = 102 -
38n_f/3$ in the perturbative expansion of $\b_V$ are the standard
ones.

\section{QCD Perturbation Theory and $\a_V$}

\subsection{BLM Scale Fixing for $\a_V$}

The calculation of the heavy quark potential at NNLO in
Ref. \cite{peter} allows the BLM procedure to be applied through NLO
in commensurate scale relations involving $\a_V$.  As a first step, we
may apply the BLM procedure to fix the renormalization scales in the
expression for $\a_V$ in terms of the conventional $\overline{MS}$
coupling.  The result is
\begin{equation} 
\frac{{\alpha}_V(Q)}{\pi} =
\frac{{\alpha}_{\overline{MS}}(Q_V^*)}{\pi} +
A_V\left(\frac{{\alpha}_{\overline{MS}}(Q_V^{**})}{\pi}\right)^2 +
B_V\left(\frac{{\alpha}_{\overline{MS}}(Q_V^{***})}{\pi}\right)^3 +
{\cdots}\; ,
\end{equation}
where
\begin{eqnarray}
A_V &=&  -\frac{2}{3}C_A \\ 
\nonumber\\
B_V &=&  \left(\frac{133}{144} - \frac{11}{4}{\zeta}_3
+ \frac{3}{8}{\pi}^2 - \frac{1}{64}{\pi}^4\right)C_A^2
+\left(- \frac{385}{192}  + \frac{11}{4}{\zeta}_3\right)C_AC_F \\
\nonumber\\
\ln(Q_V^*/Q) &=&  -\frac{5}{6} \\ 
\nonumber\\
\ln(Q_V^{**}/Q) &=& -{217\over192} + {21\over16}\zeta_3 +
\left({105\over 128}-{9\over8}\zeta_3 \right){C_F\over C_A}
\end{eqnarray}
and $C_A=N_c$.
As discussed above, we take $Q_V^{***}=Q_V^{**}$ at this order.

It is also useful to invert this, and express $\a_{\overline{MS}}$
itself in terms of $\a_V$.  In this case we obtain
\begin{equation}
\frac{{\alpha}_{\overline{MS}}(Q)}{\pi} = 
\frac{{\alpha}_V(Q_{\overline{MS}}^*)}{\pi}
+ A_{\overline{MS}}\left(
	\frac{{\alpha}_V(Q_{\overline{MS}}^{**})}{\pi}\right)^2 
+ B_{\overline{MS}}\left(
	\frac{{\alpha}_V(Q_{\overline{MS}}^{**})}{\pi}\right)^3
+ {\cdots}\; ,
\end{equation}
where
\begin{eqnarray}
A_{\overline{MS}} &=& \frac{2}{3}C_A \\ 
\nonumber\\
B_{\overline{MS}} &=&  \left(-\frac{5}{144} + \frac{11}{4}{\zeta}_3
- \frac{3}{8}{\pi}^2 + \frac{1}{64}{\pi}^4\right)C_A^2
+\left(\frac{385}{192} - \frac{11}{4}{\zeta}_3\right)C_AC_F \\
\nonumber\\
\ln(Q_{\overline{MS}}^*/Q) &=& \frac{5}{6} \\ 
\nonumber\\
\ln(Q_{\overline{MS}}^{**}/Q) &=& {103\over192} + {21\over16}\zeta_3 +
\left({105\over 128}-{9\over8}\zeta_3 \right){C_F\over C_A}\; .
\end{eqnarray}

\subsection{$e^+e^-$ Annihilation Cross Section}

We next present the NNLO scale-commensurate expansion of $\a_R$ in
terms of $\a_V$.  This is obtained by applying the BLM procedure at
NLO to the expansion of each of these observables in the $\overline
{MS}$ scheme, and then algebraically eliminating $\alpha_{\overline
{MS}}$.  The result is
\begin{equation}
\frac{{\alpha}_R(Q)}{\pi} = \frac{{\alpha}_V(Q_R^*)}{\pi} 
+ A_R\left(\frac{{\alpha}_V(Q_R^{**})}{\pi}\right)^2 
+ B_R\left(\frac{{\alpha}_V(Q_R^{**})}{\pi}\right)^3 + {\cdots}\; ,
\label{arcsr}
\end{equation}
where (for $N_c=3$)
\begin{eqnarray}
A_R &=& \frac{25}{12} \\
\nonumber\\
B_R &=& \frac{97}{72} - \frac{27}{8}{\pi}^2 + \frac{9}{64}{\pi}^4
  + \frac{10}{d(r)}\left(\frac{11}{144}-\frac{{\zeta}_3}{6}\right)
  \frac{\left({\sum}_qe_q\right)^2}{{\sum}_qe^2_q} \\
\nonumber\\
\ln(Q_R^*/Q) &=&  - \frac{23}{12} + 2{\zeta}_3 
	+ (33-2n_f)\left[-\frac{119}{864} + \frac{{\pi}^2}{72} 
	-\frac{7}{9}{\zeta}_3 + \frac{2}{3}{\zeta}_3^2\right]
	\left(\frac{{\alpha}_V(Q)}{\pi}\right) \\
\nonumber\\
\ln(Q_R^{**}/Q) &=&  -\frac{157}{60} + \frac{233}{50}{\zeta}_3 
	-2{\zeta}_5\; .
\end{eqnarray}
In Eqn. (3.13), $d(r)$ is the dimension of the quark representation,
i.e., 3 for $SU(3)$.  This relation represents an unambiguous,
fundamental test of perturbative QCD which is independent of
renormalization scale or scheme.

In order to make a comparison to experimental data, we will introduce
a parameterization of $\a_V$ which is fit to lattice data
\cite{davies} in the moderate- to high-$Q^2$ regime.  Specifically, we
take
\begin{equation}
\alpha_V(Q) = {4 \pi \over {\beta_0 \ln \left({{Q^2 + 4m_g^2}\over
\Lambda^2_V}\right)}}\; .
\label{frozencoupling}
\end{equation}
Asymptotically this reproduces the perturbative coupling, while the
effective ``gluon mass'' $m_g$ results in $\a_V$ becoming essentially
frozen for $Q^2\leq 4m_g^2$.  This form can be motivated on various
theoretical grounds {\cite{cornwall}}, and it has also been successful
in phenomenological analyses \cite{bjpr97}.

The parameters ${\L}_V$ and $m_g^2$ have been determined in Ref.
{\cite{bjpr97}}, by fitting to a lattice calculation of $V(Q^2)$
{\cite{davies}} at relatively high $Q^2$ and to a value of
${\alpha}_R$ advocated in {\cite{mattstev}}, using Eqn. (\ref{arcsr})
at LO.  They were found to be ${\L}_V=0.16~{\rm GeV}$ and $m_g^2 =
0.19~{\rm GeV^2}$.

\begin{figure}
\centerline{
\psfig{figure=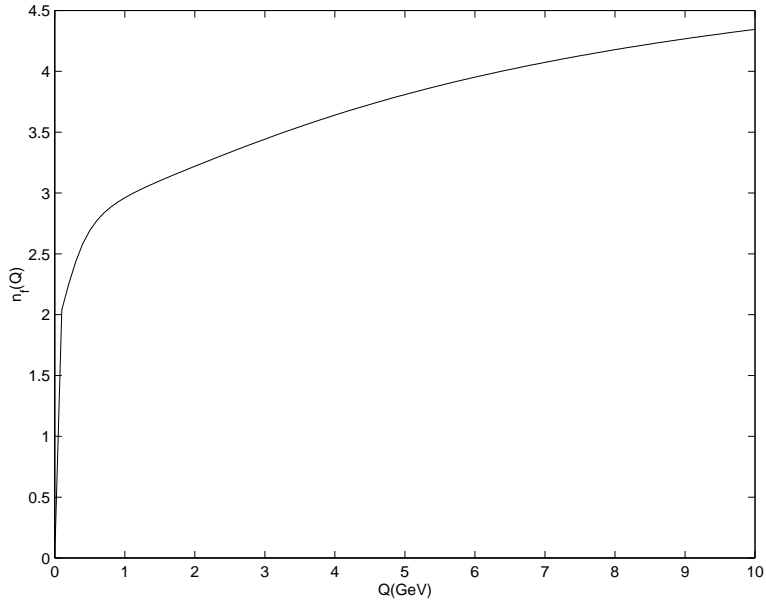,width=4.0in}
}
\vspace{.1in}
\caption{The momentum dependence of $n_f(Q^2)$.}
\label{nfonq}
\end{figure}

\begin{figure}
\centerline{
\psfig{figure=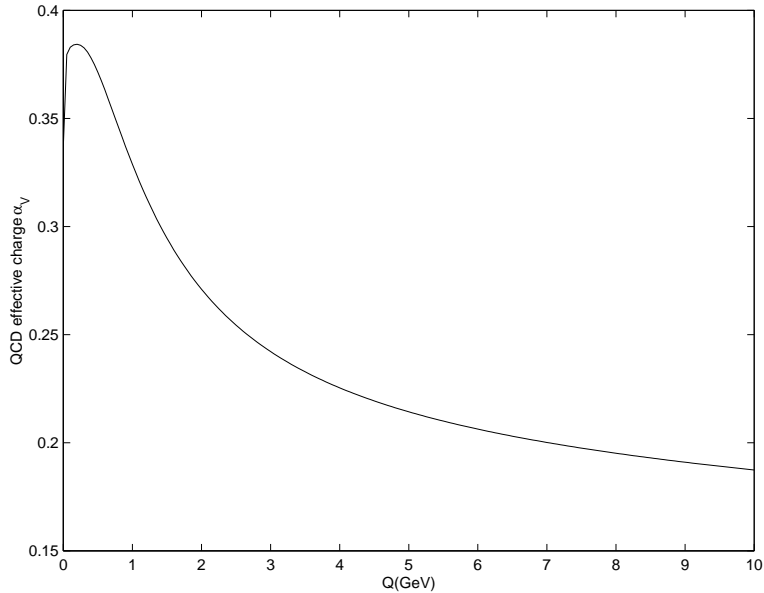,width=4.0in}
}
\vspace{.1in}
\caption{The effective charge ${\alpha}_V$, as given by Eqn. (3.16)}
\label{vsurface}
\end{figure}

Note that in the beta function ${\beta}_0$ we use a ``smeared''
function for the number of flavors, although this only affects the
low-energy regime where several quark flavor thresholds occur.  This
function is
\begin{equation}
n_f(Q^2) = \sum_f \int_0^1dz \frac{6Q^2z^2(1-z)^2} {m_f^2+Q^2z(1-z)}
\; ,
\end{equation}
and is motivated in Ref. \cite{nfsmear}.  
The integration over $z$ in Eqn. (3.17) leads to the explicit
representation\footnote{Note that
$\int_0^1 dz \frac {6z^2(1-z)^2}{x + z(1-z)} = 
1 - 6 x + {\frac{12 x^2}{(1+4x)^{1/2}}}\ln\left(\frac{(1+4x)^{1/2}+1}
{(1+4x)^{1/2}-1}\right)$.}
of the function that is identical to the $Q^2$ 
logarithmic
derivative of the one-loop massive $\beta$-function presented in Ref.
\cite{nfsmear}. 
In Fig. \ref{nfonq} we show
$n_f(Q^2)$ in the low-energy region.  We have taken $m_s=0.15~{\rm
GeV}$, $m_c=1.9~{\rm GeV}$, $m_b=4.5~{\rm GeV}$ for the quark masses.
The resulting $\a_V$ is shown in Fig. \ref{vsurface}.

Note also that for low $Q^2$ the couplings, although frozen, are
large.  Thus the NLO and higher-order terms in the CSRs are large, and
they do not give accurate results at low scales.  In addition,
higher-twist contributions to the effective charges, which are not
reflected in CSRs relating them, may be expected to be important for
low $Q^2$.
However, series expansions in terms of physical charges are likely to
be more convergent than those cast in terms of unphysical couplings
such as $\a_{\overline{MS}}$, which is singular at finite
scales.\footnote{For example, in the 't Hooft scheme
$\a_{\overline{MS}}$ has a simple pole at
$Q=\Lambda_{\overline{MS}}$.}  Thus it is quite possible that
expansions of the type we are considering can be extended to lower
physical scales than series written in terms of $\a_{\overline{MS}}$.
In any case, we will not be directly concerned with the low-$Q^2$
regime here.

Before discussing the results, it is useful to understand what
improvements we can expect from the commensurate scale relations.
First of all, of course, we have a scale-independent result, so
aesthetically we have an advantage over the conventional treatment.
Moreover, because of this we expect our result to be numerically more
accurate than previous results with the scale fixed to certain value.
The main applicability and usefulness of commensurate scale relations
is for the intermediate energy regime.  Pertubation theory is valid
only above the characteristic QCD scale ${\Lambda}_{QCD}$, and since
the commensurate scale analysis crucially depends on the validity of
perturbation theory, we don't expect much improvement in the very low
energy regime. Furthermore, in the high energy limit the residual
scale dependent terms go to zero, so scale relations are meaningless.
The $e^+e^-$ annihilation data, as well as the Bjorken sum rule data
presented in the next section, lies in the intermediate energy regime
where we expect improved predictions.

Two additional modifications of Eqn. (\ref{arcsr}) were performed
before comparing with data.  First, we have included the leading-order
electroweak corrections to account for the $Z^0$ current, which is
particularly important above 30 GeV. In addition we have included the
charm and bottom mass corrections, which are important in the range
3--15 GeV. The effect of these modifications is to replace the factor
$\sum_q{e_q^2}$ in Eqn. (\ref{alphardef}) by
\begin{eqnarray}
\sum_q\sqrt{1-\frac{4m_q^2}{Q^2}}\Biggl{[}e_q^2\left(1+\frac{2m_q^2}{Q^2}\right) 
    &+& 2{\rm Re({\it r})}e_qc_V^ec_V^q\left(1+\frac{2m_q^2}{Q^2}\right)
\\
    &+& |r|^2\left((c_V^e)^2+(c_A^e)^2\right)
	\left((c_V^q)^2\left(1+\frac{2m_q^2}{Q^2}\right)
    +(c_A^q)^2\left(1-\frac{4m_q^2}{Q^2}\right)\right)\Biggl{]}\; ,
\nonumber
\end{eqnarray} 
where
\begin{eqnarray}
c_V^q &=& I_3 - 2 {e_q} {\sin}^2 {\theta}_W \; ,
\nonumber
\\
c_V^e &=& 2 {\sin}^2 {\theta}_W - \frac{1}{2} \;,
\nonumber
\\
c_A^q &=& I_3 \;,
\nonumber
\\
c_A^e &=& -\frac{1}{2} \;,
\nonumber
\\
r &=& \frac{\sqrt{2}GM_Z^2}{Q^2-M_Z^2+iM_Z{\Gamma}_Z} 
\left(\frac{Q^2}{e^2}\right)\;
\nonumber
\\
&=& \frac{Q^2}{{\sin}^2 {2{\theta}_W} (Q^2-M_Z^2+iM_Z{\Gamma}_Z)}\;.
\end{eqnarray}
Here $I_3$ is the third component of the weak isospin of the quark
coupled to $Z^0$ and the weak mixing angle $\theta_W$ is given by
${\sin}^2 {\theta}_W = 0.22$. The mass and the decay width of $Z^0$
are given by $M_Z = 91.2~\rm{GeV}$ and ${\Gamma}_Z = 2.5~\rm{GeV}$,
respectively.
\begin{figure}
\centerline{
\psfig{figure=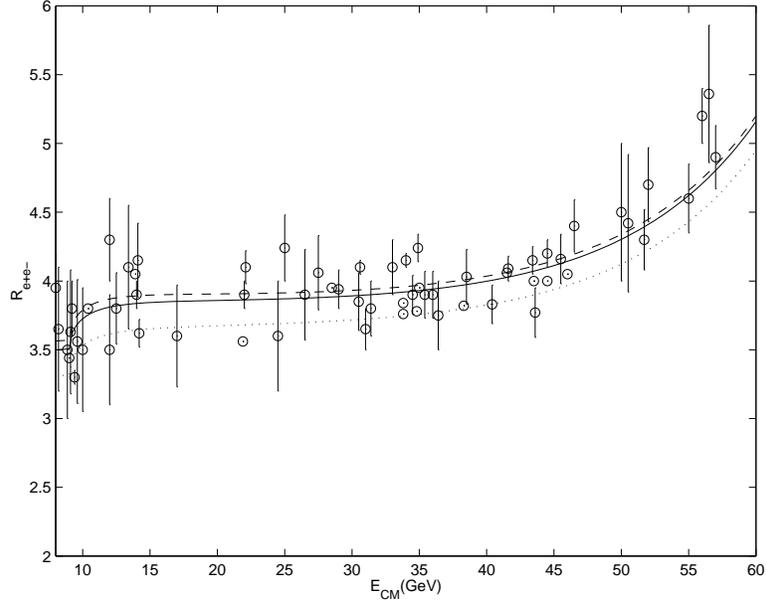,width=4.0in}
}
\vspace{.1in}
\caption{The scale-commensurate expansion of $\a_R$ in terms of $\a_V$
in the high energy regime. The solid line is given by Eqn. (3.11); the
dashed line is the prediction quoted by the PDG; the dotted line is
the leading order result (with mass and electroweak corrections). }
\label{plothigh}
\end{figure}

\begin{figure}
\centerline{
\psfig{figure=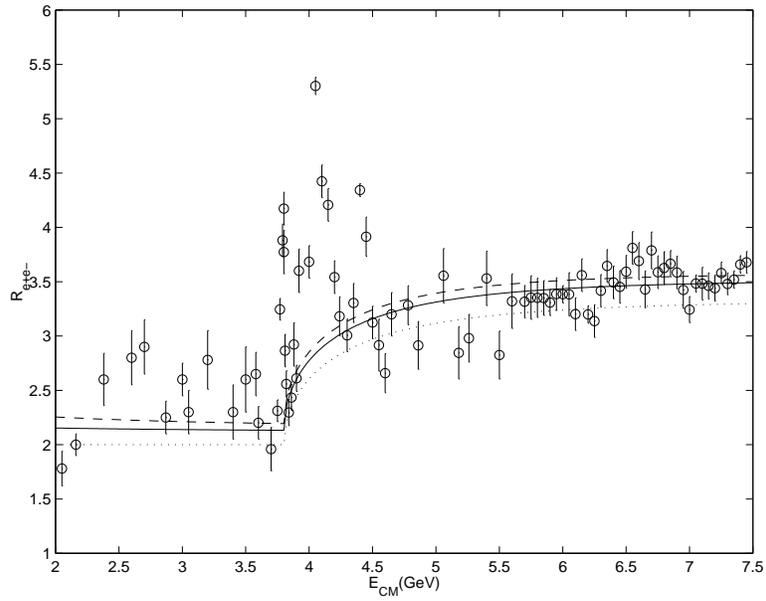,width=4.0in}
}
\vspace{.1in}
\caption{Scale-commensurate expansion of $\a_R$ in terms of $\a_V$ in
the intermediate-energy regime.}
\label{plotlow}
\end{figure}

In Fig. \ref{plothigh}, we show the commensurate scale result
(\ref{arcsr}) along with a representative subset of the available data
{\cite{pdg}} in the energy range 8--60 GeV.
We find our results to be in excellent agreement with the data, as
well as the standard QCD predictions quoted by the Particle Data Group
{\cite{DS}} with the scale fixed to a certain value
($\L_{\overline{MS}}=0.25 \rm{GeV}$).  In Fig. \ref{plotlow}, we show
our theoretical prediction and the data in the 2--7.5 GeV range.
Again, we find very nice agreement with the data, particularly
considering that we have neglected corrections from the
$J/{\psi}(1S)$, ${\psi}(2S)$, and other vector meson resonances.
Note that the data for $3.6~\rm{GeV}<Q<7.5~\rm{GeV}$ has been
subtracted by
$0.84\sqrt{1-{4m_\tau^2}/{Q^2}}\left(1+{2m_\tau^2}/{Q^2}\right)$ to
account for hadronic production that proceeds via tau lepton pairs,
which the early experiments did not distinguish from quark-hadron
processes. The factor $0.84=1-(2/5)^2$ is the probability that either
tau will decay to hadrons.

\subsection{Bjorken Sum Rule}

\begin{figure}
\centerline{
\psfig{figure=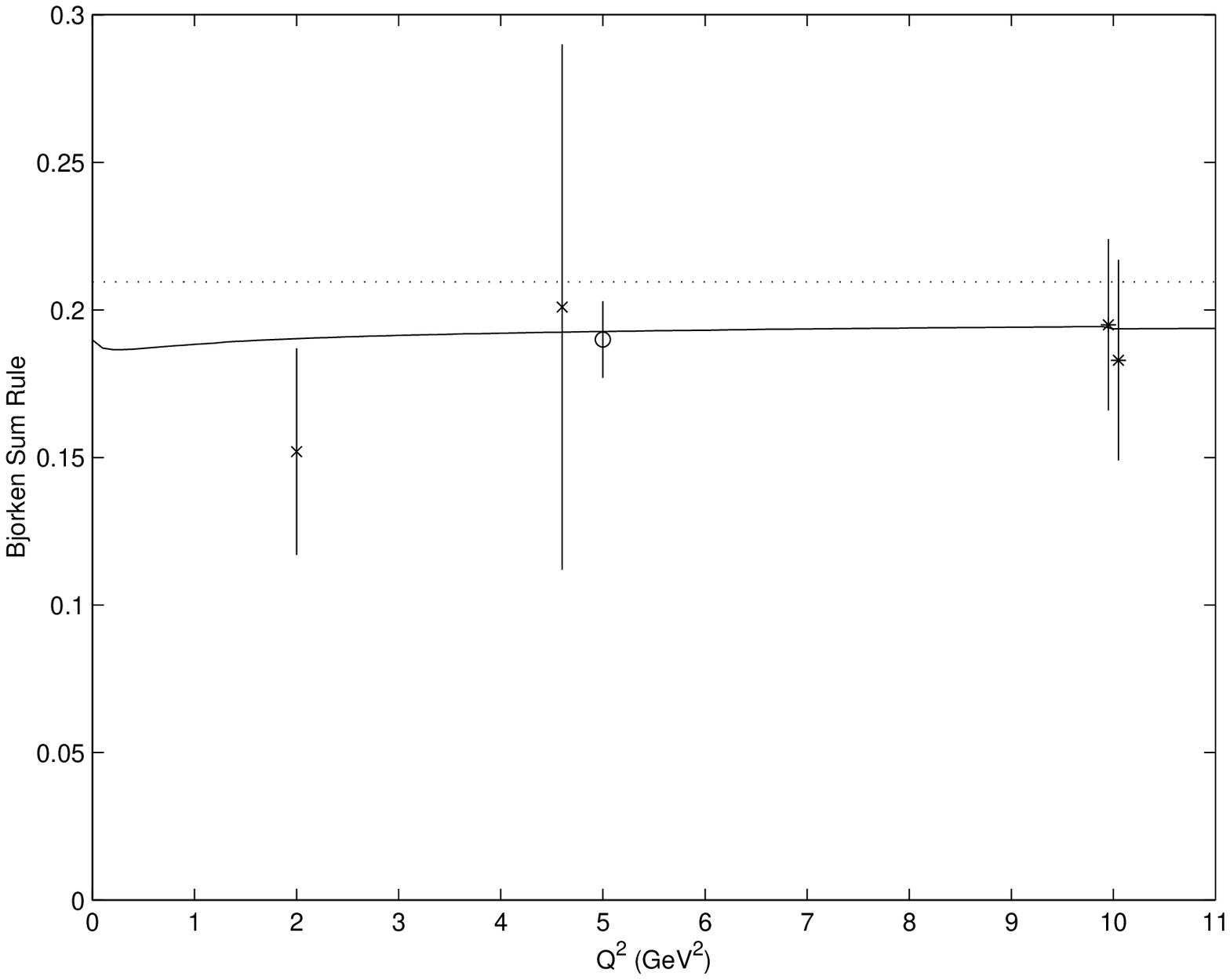,width=4.0in}
}
\vspace{.1in}
\caption{The dotted line shows the leading-order prediction for the
Bjorken sum rule while the solid line includes the scale-commensurate
corrections through NNLO.  Also shown are data from Ref. [18]
(crosses), from the E154 Collaboration [19] (circles), and from the
SMC Collaboration [20] (stars).}
\label{bsr}
\end{figure}

Finally we present the scale-commensurate expansion of the Bjorken sum
rule in terms of $\a_V$ at NNLO. The result is
\begin{equation}
\frac{{\alpha}_{g_1}(Q)}{\pi} = \frac{{\alpha}_V(Q_g^*)}{\pi} 
+ A_g\left(\frac{{\alpha}_V(Q_g^{**})}{\pi}\right)^2 
+ B_g\left(\frac{{\alpha}_V(Q_g^{**})}{\pi}\right)^3 + {\cdots}\; ,
\label{agcsr}
\end{equation}
where
\begin{eqnarray}
A_g &=& \frac{13}{12}\\
\nonumber\\
B_g &=& -\frac{131}{72} - \frac{27}{8}{\pi}^2 + \frac{9}{64}{\pi}^4\\
\nonumber\\
\ln(Q_g^*/Q) &=&  - \frac{1}{6}
-\frac{43}{144}\left(11-\frac{2}{3}n_f\right)
\left(\frac{{\alpha}_V(Q)}{\pi}\right)\\
\nonumber\\
\ln(Q_g^{**}/Q) &=& -\frac{191}{117} - \frac{5}{78}{\zeta}_3 
+ \frac{30}{13}{\zeta}_5\; .
\end{eqnarray}

In Fig. \ref{bsr} we show the commensurate scale result to NNLO and
the leading order perturbative result with the five currently
available data points.  This plot strongly suggests that the higher
order pQCD corrections do indeed give the correct convergence to the
physical result. Our results may also be compared with an analysis of
the Bjorken sum rule {\cite{aptbsr}} using so-called analytic
perturbation theory (APT) {\cite{apt}}. In Ref. {\cite{aptbsr}}, the
authors show that by requiring the QCD coupling ${\alpha}_s$ to be
analytic, thereby removing unphysical singularities, they can obtain
approximately scheme independent results. Their plot of the correction
to the Bjorken sum rule, $\alpha_{g_1}/\pi$, is very similar to what
we obtain using commensurate scale relations.

\section{Conclusions}

In this paper, we have applied the physical QCD effective charge
${\alpha}_V$, defined by the heavy quark potential, in calculations of
the $e^+e^-$ annihilation cross section and the Bjorken sum rule.
Following the BLM procedure, we derived the NNLO scale-commensurate
expansions of ${\alpha}_R$ and ${\alpha}_{g_1}$ in terms of
${\alpha}_V$ and used these expansions to numerically compute the
$e^+e^-$ annihilation cross section and the Bjorken sum rule.  Using a
phenomenological form for the effective charge ${\alpha}_V$
[Eqn. (\ref{frozencoupling})] which is consistent with the lattice
determination of the heavy quark potential, we obtain excellent
agreement between our results and the experimental data in both cases.
Furthermore, because of the scale independence, we trust that our
results are numerically more accurate than previous results with the
scale fixed to a certain value.  The application of scale-commensurate
expansions to other observables is forthcoming.

\acknowledgements 
\noindent
This work was supported in part by a grant from the US Department of
Energy.

\end{document}